\documentclass[
 reprint,twocolumn,superscriptaddress,
 amsmath,amssymb,
 aps,showpacs,
]{revtex4}

\usepackage{framed}
\usepackage{graphicx}
\usepackage{dcolumn}
\usepackage{bm}
\usepackage[titletoc,title]{appendix}
\usepackage{color}
\DeclareMathAlphabet{\mathbbold}{U}{bbold}{m}{n}

\begin{document}


\author{Jirawat Tangpanitanon}
\email{a0122902@u.nus.edu}
\affiliation{Centre for Quantum Technologies, National University of Singapore, 3 Science Drive 2, Singapore 117543}
\author{Stephen R. Clark}
\affiliation{Department of Physics, University of Bath, Claverton Down, Bath BA2 7AY, United Kingdom}
\affiliation{Max Planck Institute for the Structure and Dynamics of Matter, CFEL, Hamburg, Germany}
\author{V. M. Bastidas}
\affiliation{NTT Basic Research Laboratories $\&$ Research Center for Theoretical Quantum Physics, 3-1 Morinosato-Wakamiya, Atsugi, Kanagawa, 243-0198, Japan}
\author{Rosario Fazio}
\affiliation{Abdus Salam ICTP, Strada Costiera 11, I-34151 Trieste, Italy}
\affiliation{NEST, Scuola Normale Superiore and Istituto Nanoscienze-CNR, I-56126 Pisa, Italy}
\author{Dieter Jaksch}
\affiliation{Clarendon Laboratory, University of Oxford, Parks Road, Oxford OX1 3PU, United Kingdom}
\affiliation{Centre for Quantum Technologies, National University of Singapore, 3 Science Drive 2, Singapore 117543}
\author{Dimitris G. Angelakis}
\email{dimitris.angelakis@qubit.org}
\affiliation{Centre for Quantum Technologies, National University of Singapore, 3 Science Drive 2, Singapore 117543}
\affiliation{School of Electrical and Computer Engineering, Technical University of Crete, Chania, Greece 73100}


\title{Hidden Order in Quantum Many-body Dynamics of Driven-Dissipative \\ Nonlinear Photonic Lattices}

\date{\today}

\begin{abstract}
We study the dynamics of nonlinear photonic lattices driven by two-photon parametric processes. By means of matrix-product-state based calculations, we show that a quantum many-body state with long-range hidden order can be generated from the vacuum. This order resembles that characterizing the Haldane insulator. A possible explanation highlighting the role of the symmetry of the drive, and the effect of photon loss are discussed. An implementation based in superconducting circuits is proposed and analyzed \end{abstract}


\maketitle
\section{Introduction} 
Advances in quantum optics over the past decades have made it possible to engineer strong interactions between individual photons \cite{lukin2014}. This motivates its use for generating new kinds of strongly correlated states of light and matter \cite{dimitrisreview,dimitrisbook2017} for quantum simulation \cite{zoller2012,Johnson2014}. Indeed, early theoretical works have shown that arrays of coupled nonlinear cavities can exhibit a Mott insulator to superfluid phase transition of light, if dissipation is negligible \cite{dimitris2007,hartmann2006,greentree2006}. Subsequent works have also shown the possibility to realize a family of many-body phenomena with photons including effective spin models \cite{0295-5075-84-2-20001}, the fractional quantum hall effect \cite{PhysRevLett.101.246809}, and topologically protected transport of quantum states \cite{PhysRevLett.117.213603}. Moreover, the signatures of localization of interacting photons in a quasi-periodic potential have recently been observed with a nine-site superconducting circuit \cite{Roushan1175}. 

Interacting photons provide a natural setting for simulating open quantum systems because photons dissipate to the environment and because they can be coherently driven. The coupling to the environment is usually assumed to be weak and the bath is memoryless in which case the system could reach a dynamically-stable steady state that depends on the symmetries of the system \cite{PhysRevA.89.022118,dimitris2012}. Early theoretical works have shown that such steady states manifest various quantum many-body phases \cite{carusotto2009, hartmann2010, carusotto2012, bardyn2012, jin2013, PhysRevA.87.053846,houck2016,PhysRevA.93.023821} and can exhibit a dissipative phase transition (DPT) \cite{eisert2017, PhysRevA.86.012116}. A nonlinear superconducting circuit with up to 72 sites has also been fabricated to study DPT \cite{houck2016}.

Following the success of Landau's symmetry breaking theory in describing classical and ground-state phases of matter, local order parameters have also been used to classify these new non-equilibrium steady-state phases \cite{eisert2017, PhysRevA.86.012116,carusotto2009, hartmann2010, carusotto2012, dimitris2012, bardyn2012, jin2013, houck2016}. However, in equilibrium systems, there are phases that do not follow Landau's symmetry breaking theory \cite{wen2017}. The latter can be probed by, for example, non-local order parameters \cite{cirac2008, schuch2012} or the existence of edge states \cite{kane2010,choucheng2011}. These phases are symmetry protected topological (SPT) phases \cite{xiaogang2009,xiaogang2010} and phases with topological orders \cite{wen1990, nayak2008}. Experimental realizations of topological phases have been explored in various quantum technology platforms including cold atoms \cite{zoller2016} and photonic systems \cite{2018arXiv180204173O,marin}.

In this work, we study the role of a non-local order parameter in the driven-dissipative dynamics of a quantum many-body system and its connection to the underlying symmetry. Specifically, we consider a non-local hidden order, analogous to the famous SPT phase characterizing the equilibrium Haldane insulator (HI) phase \cite{torre2006, berg2008, rossini2012}. The system we consider is a lossy nonlinear photonic lattice of the extended Bose-Hubbard type \cite{torre2006, berg2008, rossini2012} which in the right regime can be mapped to the spin-1 Haldane model \cite{kennedy1992, oshikawa2012} and driven by a two-photon parametric process \cite{bardyn2012, jonathan2013, devoret2013}. Using matrix-product-state based calculations \cite{mps, tnt2016}, we show that this process drives the vacuum into a quantum many-body state with non-zero hidden order. We argue that this effect is due to the symmetry of the parametric drive, which cannot be achieved by a conventional one-photon coherent drive. We analyze this symmetry analytically and numerically by showing that the parametric drive respects the symmetry of the HI state. However, single photon losses break this symmetry and eventually destroy the hidden order in the steady state. We note that the decoherence process due to losses can be suppressed by engineering the environment to drive the system into the desired state \cite{cirac2009, zoller2008, lukin2017}.

This paper is organized as follows. We describe our system including the definition of the hidden order in Sec. \ref{sec:the_system}. The symmetry of the parametric process is analyzed in Sec. \ref{sec:symmetry_analysis}. Numerical simulations of the driven-dissipative dynamics showing the evolution of the hidden order are shown and discussed in Sec. \ref{sec:main_result}. In Sec. \ref{sec:onsite}, we discuss a conventional one-photon coherent drive which breaks the symmetry of the Haldane phase. We conclude in Sec. \ref{sec:conclusion}.


\begin{figure}
    \includegraphics[width=8.5cm,height=6cm]{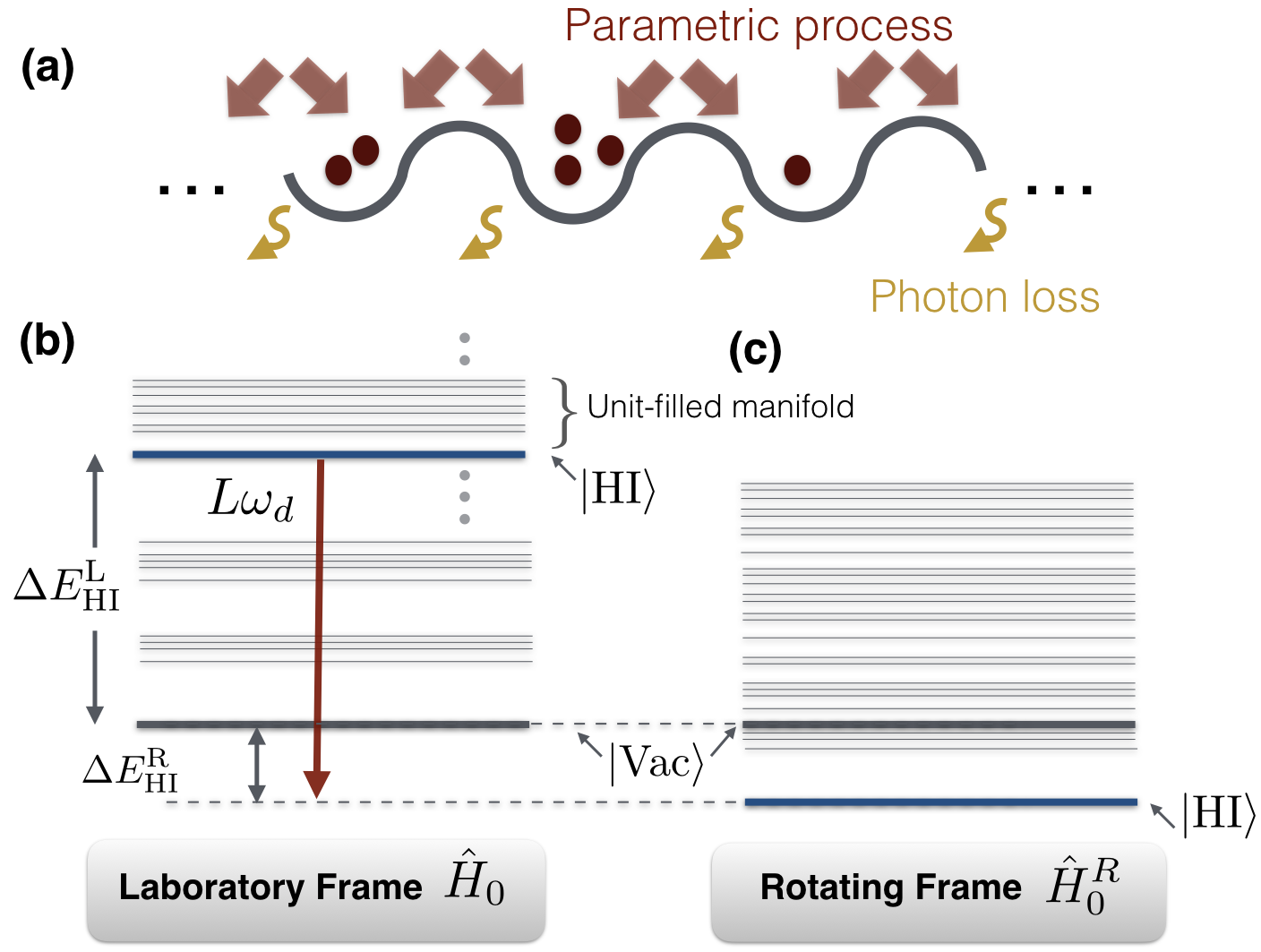}
    \caption{ (a) Sketch of the one-dimensional lossy photonic lattice described by the EBH model and driven by parametric drive. Lower panels are energy spectra of the undriven Hamiltonian in Eq.\ref{eq:ebh}. (b) the laboratory frame and (c) the rotating frame.The red arrow indicates the energy shift $L\omega_d$ of the HI state due to the rotating frame. Since the HI state is a unit-filled state, its energy will be lowered by $L\omega_d$.}
    \label{fig1}
\end{figure}

\section{The system}
\label{sec:the_system}
We consider a 1D coupled nonlinear resonator array described by the Hamiltonian $\hat{H}_{\text{tot}}^{\rm Lab}=\hat{H}_0+\hat{H}_{\rm drv}^{\rm par}$, where $\hat{H}_0$ is the extended Bose-Hubbard (EBH) model ($\hbar =1$), 
\begin{align}
\hat{H}_0^{\rm Lab} = &  \omega_r\sum_{i=1}^L \hat{n}_i -J\sum_{i=1}^{L-1}\left(\hat{a}^\dagger_i\hat{a}_{i+1}+H.c.\right) \nonumber \\
 		   &+\frac{U}{2}\sum_{i=1}^L \hat{n}_i(\hat{n}_i-1) +V\sum_{i=1}^{L-1}\hat{n}_i\hat{n}_{i+1},
\label{eq:ebh}
\end{align}
where $\omega_r$ is the frequency of the resonator, $J$ is the hopping strength, $U$ is the on-site Kerr nonlinearity, and $V$ is the cross-Kerr nonlinearity, and $L$ is the number of sites, see Fig. \ref{fig1}(a). The operator $\hat{a}_i$ is a bosonic annihilation operator at site $i$ and $\hat{n}_i=\hat{a}^\dagger_i\hat{a}_i$ is a local number operator, respectively. The bosonic operators obey the commutation relation $\left[\hat{a}_i,\hat{a}_j^\dagger\right]=\delta_{i,j}$ and $\left[\hat{a}_i,\hat{a}_j\right]=0$. Throughout the paper we will consider the regime $U\gg J$ which allows us to map the bosonic system onto a spin-1 chain as detailed below. The system is subjected to two-photon nearest-neighbor parametric driving \cite{bardyn2012,jonathan2013,devoret2013}
\begin{equation}
\hat{H}_{\rm drv}^{\text par}=\Omega\sum_{i=1}^{L-1}(\hat{a}_i\hat{a}_{i+1}e^{2i\omega_dt}+H.c.),
\label{eq:}
\end{equation}
where  $\omega_d$ is the driving frequency and $\Omega$ is the amplitude of the drive. A detailed discussion on circuit-QED implementations of this Hamiltonian including parametric drives is presented in Appendix B. We remove the time dependence of the drive by going to the rotating frame defined by $\hat{R}=\exp(i \omega_d t\sum_{i=1}^L\hat{n}_i)$. The new Hamiltonian is 
\begin{align}
\hat{H}_{\text{tot}}^{\rm R}=&\hat{R}\hat{H}^{\rm Lab}_{\text{tot}}\hat{R}^\dagger-i\hat{R}\partial_t \hat{R}^\dagger \nonumber \\=&\hat{H}^{\rm R}_0+\hat{H}^{\rm R}_{\rm drv},
\end{align}
where
\begin{align}
\hat{H}^{\rm R}_0=&-\mu\sum_{i=1}^L \hat{n}_i -J\sum_{i=1}^{L-1}\left(\hat{a}^\dagger_i\hat{a}_{i+1}+H.c.\right) \nonumber \\
 		   &+\frac{U}{2}\sum_{i=1}^L \hat{n}_i(\hat{n}_i-1) +V\sum_{i=1}^{L-1}\hat{n}_i\hat{n}_{i+1},  \\
\hat{H}^{\rm R}_{\rm drv} =& \Omega\sum_{i=1}^{L-1}(\hat{a}_i\hat{a}_{i+1}+H.c.),
\end{align}
and $\mu = \omega_d-\omega_r$ is the detuning. In the following discussion, we will analyze the properties of $\hat{H}_{\text{tot}}^R$ both when $\Omega=0$ and $\Omega>0$.The driven-dissipative dynamics is governed by the Master equation,
\begin{equation}
\frac{\partial}{\partial t} \hat{\rho}=-i[\hat{H}_{\rm tot}^{\rm R},\hat{\rho}]  -\frac{\gamma}{2}\sum_{i=1}^L(\{\hat{n}_i,\hat{\rho}\}-2\hat{a}_i\hat{\rho} \hat{a}^\dagger_i),
\label{eq:lms}
\end{equation}
where $\gamma$ is the dissipation rate, $\rho$ is the density matrix of the system.

We study the quantum phase via hidden order defined by a non-vanishing string order (SO)
\begin{equation}
\mathcal{O}_{\rm S}=\lim _{|i-j|\to \infty}|\langle \delta \hat{n}_i e^{i\pi \sum^{j-1}_{k=i+1} \delta\hat{n}_k} \delta\hat{n}_j\rangle|>0,
\end{equation}
and a vanishing density-wave order (DWO)
\begin{equation}
\mathcal{O}_{\rm DW}=\lim _{|i-j|\to \infty} |\langle \delta \hat{n}_i  \delta \hat{n}_j\rangle |=0,
\end{equation}
where $\delta \hat{n}_i=\hat{n}_i-\bar{n}$ is the number fluctuation at site $i$ and $\bar{n}=\sum_{i=1}^{L}\langle\hat{n}_i\rangle/L$ is the filling factor \cite{torre2006}. The vanishing DWO implies that quantum fluctuations between two distant sizes are uncorrelated. Yet the non-vanishing SO implies that these fluctuations exhibit a certain infinitely long-range structure which is `hidden' from DWO. Note that the string order operator is not hermitian, hence SO is not a correlation function.This hidden order is used to characterize the topological Haldane phase with unit filling in the equilibrium context Appendix A. Non-equilibrium quench dynamics and thermalization of SO in the context of the spin-chain system have been studied in Ref. \cite{fazio2014,fazio2016}. In Ref. \cite{fazio2014,fazio2016}, the authors assume that the starting state already has SO. In contrast, here we show in Sec. \ref{sec:main_result} that SO can be generated from the vacuum in the driven-dissipative senario.


\section{Symmetry of the two-photon parametric process}
\label{sec:symmetry_analysis}

In this section, we will analyze the symmetry of the two-photon parametric process by mapping the bosonic system into a spin-chain system. Then we analytically and numerically show that SO of the Haldane phase is robust against weak parametric driving.

We first examine the energy spectrum of the EBH model in the context of the coupled resonator array, ignoring dissipation. The EBH model conserves the number particles, hence the excited states can be grouped into manifolds labelled by the total number of particles $N$ which is an eigenvalue of $\sum_i\hat{n}_i$. Since we work in a regime far from the ultra-strong coupling regime, i.e. $\omega_r\gg J, U,V$, the ground state of the undriven system is the vacuum, see Fig. \ref{fig1}(b). It has been shown that, at appropriate parameter regimes, the lowest energy state in the unit-filled manifold ($N=L$) shows the topological Haldane insulator (HI) phase, exhibiting the hidden order \cite{torre2006, berg2008, rossini2012}. We label the many-body state in this phase as $|\rm HI\rangle$.

As will be shown below, the detuning $\mu$ can be chosen such that $|\rm HI\rangle$ becomes a gapped ground state of $\hat{H}^{\rm R}_{\rm 0}$, see Fig. \ref{fig1}(c). We consider a weak drive $\Omega< U,V,J$ such that the filling factor of $|\rm HI\rangle$ is approximately unaffected by the drive due to the gap. We numerically confirm that this approximation is valid below. We then map the bosonic system onto a spin-1 chain model by only keeping states with site occupation of up to 2 photons. This is justified by the large on-site interaction $U\gg J$ required for the insulating phases. As a result, the bosonic Fock states  $\{|0\rangle_f, |1\rangle_f, |2\rangle_f\}$ can replaced by the spin-1 states $\{|+\rangle_s, |0\rangle_s, |-\rangle_s\}$. The bosonic operators can be replaced with spin-1 operators, i.e., $\hat{a}_i\to \hat{S}^+_i/\sqrt{2}$ and $\hat{n}_i\to \hat{\mathbbold{1}} -\hat{S}^z_j$. In the spin-chain picture, the total bosonic Hamiltonian $\hat{H}_{\text{tot}}^{\rm R}$ becomes
\begin{align}
\hat{H}^{S}_{\rm tot, par}= &\sum_{i=1}^{L-1}\left((J+\Omega^{})\hat{S}^x_i\hat{S}^x_{i+1}+(J-\Omega)\hat{S}^y_i\hat{S}^y_{i+1} \right) \nonumber \\
                           &+V \sum_{i=1}^{L-1}\hat{S}^z_i\hat{S}^z_{i+1}+\frac{U}{2}\sum_{i=1}^L(\hat{S}^z_i)^2.
\label{eq:hs}
\end{align}
An additional term $(-\mu+U/2+V)\sum_i\hat{S}^z_i$ has been dropped, as it is approximately zero since we assume that the ground state of the undriven system has unit filling and the drive is weak. The system has the global $D_2=Z_2\times Z_2$ symmetry, i.e. $\pi-$rotation of all spins about $X$, $Y$, and $Z$ axes. One can see that the presence of the weak parametric process does not alter this symmetry.

\begin{figure}
\centering
\includegraphics[width=8.5cm, height=3.5cm]{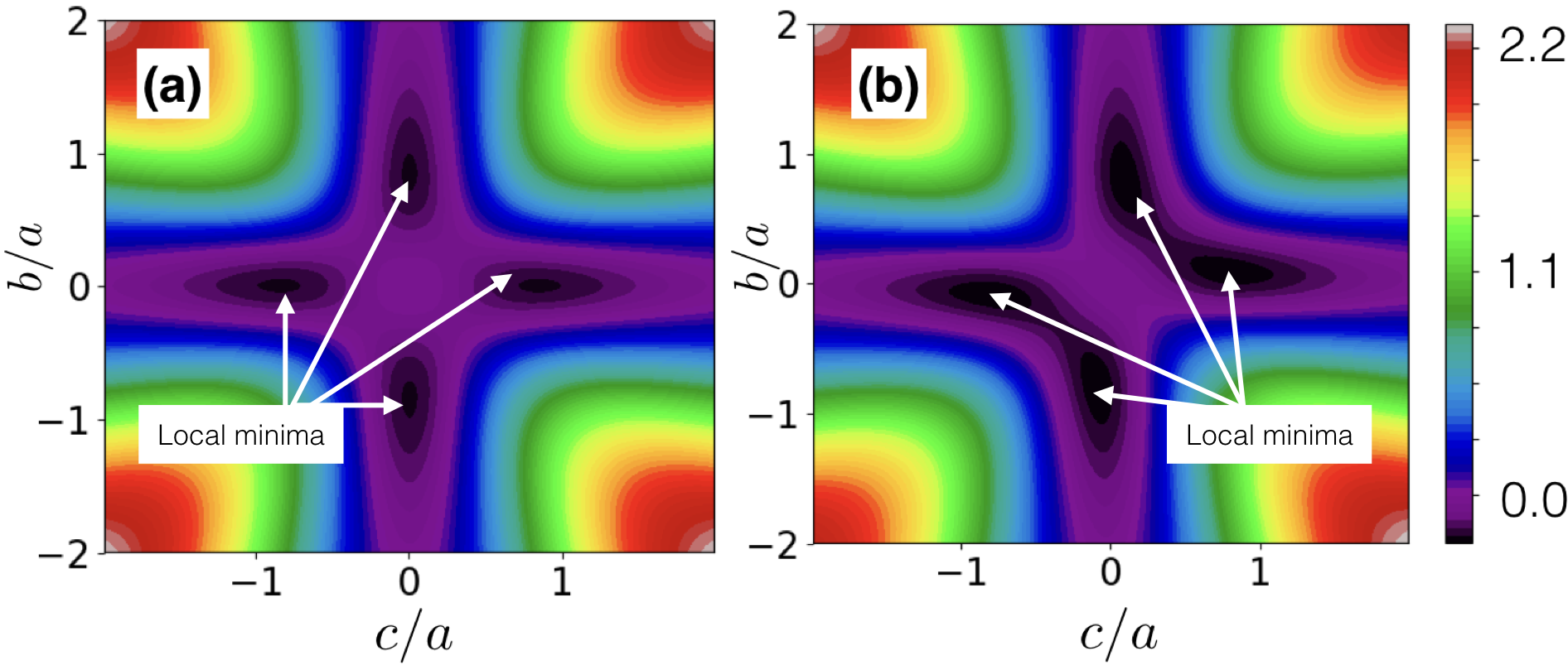}
\caption{Mean-field energy landscape $E_{MF}$ plotted against $b/a$ and $c/a$ with $V=2.8J$ and $U=5J$. (a) In the absence of drive $\Omega=0$,  the variational ferromagnetic ground state has 4-fold ground-state degeneracy, reflecting the global $D_2$ symmetry of the HI phase. (b) With a weak parametric drive $\Omega=0.1J$, the ground states remain 4-fold degenerate. Hence the symmetry is unbroken.}
\label{fig2}
\end{figure}
\begin{figure}
\includegraphics[width=8cm, height=6cm]{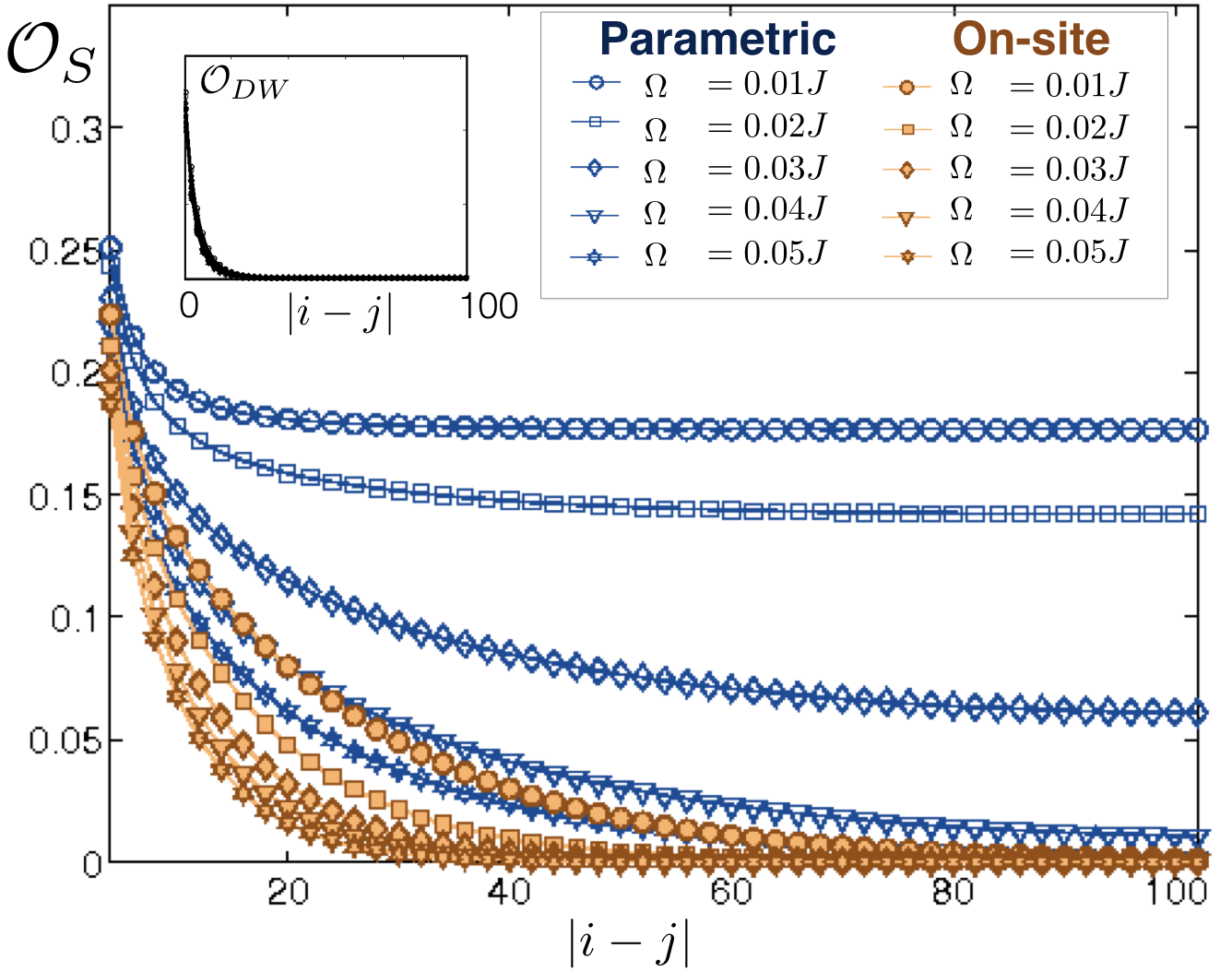}
\caption{DMRG calculations of SO and DWO of the ground state of $\hat{H}_{\rm tot}^{\rm R}$ with $\mu=7.5J$, $U=5J$ and $V=3.3J$. It shows that weak parametric drives preserve the HI phase which is characterized by a non-vanishing SO and a vanishing DWO. Both SO and DWO are zero when on-site drives are used, indicating that the HI phase is destroyed. The DMRG calculations were performed with open boundary conditions and the bond dimension of 200. The system's size is $L=300$. The local Hilbert space in the numerics is truncated at the four photon Fock state. ($\Delta E^R_{\text{HI}}=\langle \text{HI} |\hat{H}_{0}^R|\text{HI}\rangle \approx -1.22LJ$ and $\Delta E^L_{\text{HI}}=\langle \text{HI} |\hat{H}_{0}|\text{HI}\rangle \approx 6.28LJ$).}
\label{fig3}
\end{figure}

\begin{figure*}
\includegraphics[width=18cm, height=6cm]{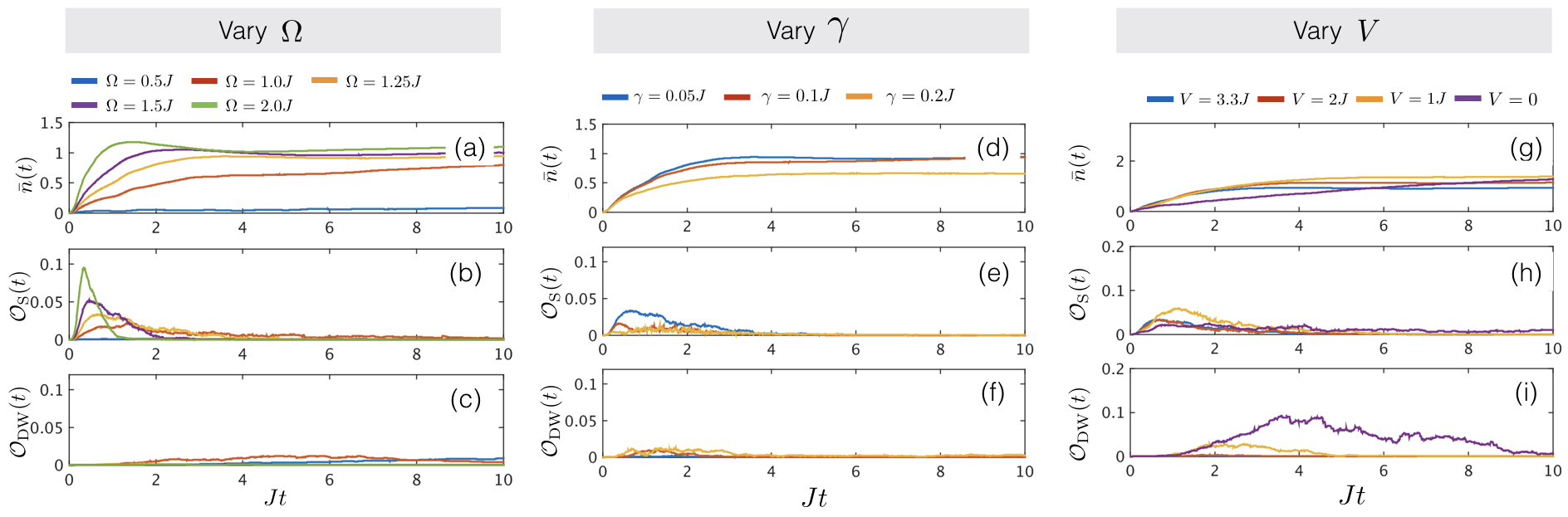}
\caption{\textbf{Driven-dissipative quantum many-body dynamics}. Time evolution of the system ($L=50$) evolving under Eq. (\ref{eq:lms}) with parametric drive, obtained using the quantum trajectories with 100 trajectories and the TEBD algorithm with the bond dimension of 100. In (a)-(c), we vary the amplitude of the drive with  $\mu=6.28J$, $U=5J$, $V=3.3J$, and $\gamma=0.05J$. In (d)-(f), we vary the photon loss rate with  $\mu=6.28J$, $U=5J$, $V=3.3J$, and $\Omega=1.25J$. In (g)-(i), we vary $V$ with  $\mu=6.28J$, $U=5J$, $\gamma=0.05J$, and $\Omega=1.25J$. }
\label{fig4} 
\end{figure*}

To understand how symmetry breaking perturbations affect the HI state using a mean-field approximation, let us consider the unitary transformation  \cite{kennedy1992}
\begin{equation}
\hat{U}_{KT}=\prod_{i<j}\exp(i\pi \hat{S}^z_i\hat{S}^x_j).
\end{equation}
This transformation is defined such that the non-local SO will be transformed into local order so that the mean-field approximation can be applied.The transformed Hamiltonian is
\begin{align}
\hat{U}_{KT} & \hat{H}^S_{\rm tot, par} \hat{U}_{KT}^{-1}  \nonumber \\ =&
-(J+\Omega)\sum_{i=1}^{L-1}\hat{S}^x_i\hat{S}^x_{i+1}  \nonumber \\ 
&- (J-\Omega)\sum_{i=1}^{L-1}\hat{S}^y_i\exp\left(i\pi (\hat{S}^z_i+\hat{S}^x_{i+1})\right)\hat{S}^y_{i+1} \nonumber \\
& -V\sum_{i=1}^{L-1}\hat{S}^z_i\hat{S}^z_{i+1}+\frac{U}{2}\sum_{i=1}^L(\hat{S}^z_i)^2.
\end{align}

We see that this Hamiltonian still involves only nearest-neighbor terms even though the transformation is non-local. This is due to the global $D_2$ symmetry of the Hamiltonian. Local terms that break the global $D_2$ symmetry will be transformed into non-local terms in this picture. As shown in Ref. \cite{kennedy1992}, the motivation for using $\hat{U}_{KT}$ is that SO in the original picture will be transformed into ferromagnetic order  (FMO), i.e., 
\begin{equation}
\hat{U}_{KT}\left(\hat{S}^z_i e^{i\pi \sum^{j-1}_{k=i+1} \hat{S}^z_k} \hat{S}^z_j\right)\hat{U}_{KT}^{-1}=\hat{S}^z_i\hat{S}^z_j.
\end{equation}  
Below we show that FMO is stable against weak parametric driving using a simple mean-field analysis as in Ref. \cite{kennedy1992} which is then backed up quantitatively by density matrix renormalization group (DMRG) calculations \cite{schollwoeck2011}.

The mean-field energy is defined as $E_{\rm MF}\equiv\langle \Phi|\hat{U}_{KT}\hat{H}^S_{\rm tot,}\hat{U}_{KT}^{-1}|\Phi\rangle$, where $|\Phi\rangle=\otimes_iA_i|\phi\rangle_i$ is a homogenous product state ansatz, $|\phi\rangle_i=a|0\rangle_i + b|+\rangle_i + c|-\rangle_i$ is a local state, $A_i=1/\sqrt{|a|^2+|b|^2+|c|^2}$ is a normalization factor, and $a,b,c$ are complex numbers. The mean-field energy takes the form
\begin{align}
E_{\rm MF} & =  \{|a|^2+|b|^2+|c|^2\}^{-2}[(\frac{U}{2}-V) (|b|^4+|c|^4) \nonumber \\
    &+ 2(\frac{U}{2}+V)|b|^2|c|^2 + (\frac{U}{2}-2J)|a|^2(|b|^2+|c|^2)\nonumber \\
    & - \operatorname{Re} \{2Ja^2(b^{*2}+c^{*2})+4\Omega|a|^2b(c+c^*)\}]
\label{eq:mf}
\end{align}
where $\bar{\cdot}$ indicates a complex conjugate. $E_{\rm MF}$ is minimized when the last two terms inside $\operatorname{Re}\{...\}$ are maximized. This happens when when $a,b$ and $c$ are real. Without loss of generality, we can set $a=1$ and get
\begin{align}
E_{\rm MF} = &\{1+b^2+c^2\}^{-2}[(\frac{U}{2}-V) (b^4+c^4)+ 2(\frac{U}{2}+V)b^2c^2 \nonumber \\
               &  + (\frac{U}{2}-4J)(b^2+c^2) - 8\Omega bc ].
\label{eq:Eabc}
\end{align}
The mean-field energy landscape is shown in Fig. \ref{fig2} with $U=5J$ and $V=2.8J$. When $\Omega=0$, $E_{\rm MF}$ displays four degenerate FM ground states. This reflects the global $D_2$ symmetry of the corresponding HI phase because the global $D_2$ symmetry implies that the state is invariant under the global $\pi-$rotation about $x,y,z$ axes. Nevertheless, $\pi-$rotation around $x$ and $y$ axes also implies $\pi-$rotation around $z$ axis. Hence the degeneracy is 4-fold. With weak parametric drive $\Omega=0.1J$, the FM ground states remain four-fold degenerate, confirming that the $D_2$ symmetry is unbroken. 

To corroborate the mean-field picture quantitatively we performed DMRG calculations on the bosonic Hamiltonian $\hat{H}_{\rm tot}^{\rm R}$ in the rotating frame. The degenerate ground states of the HI phase are lifted by forcing the edge sites to have no photon at one end and two photons at another end. The HI state was found as the ground state of $\hat{H}_{\rm 0}^{\rm R}$ by numerically scanning $\mu$. When $\Omega>0$, the SO and the DWO of the ground state for different $\Omega$ are shown in Fig. \ref{fig3}. It confirms that the HI phase is stable against weak parametric driving.


\section{The emergence of the hidden order}
\label{sec:main_result}

In this section, we turn into the driven-dissipative scenario which involves photon loss. The dynamics of the system is now described by the Lindblad master equation Eq. \ref{eq:lms}. Time evolution is obtained by solving the Lindblad Master equation (Eq. \ref{eq:lms}) using the quantum trajectories \cite{daley2014} and the Time-Evolving Block Decimation (TEBD) algorithm \cite{vidal2003}. We start from the vacuum and switch on suddenly the parametric drive.

In Fig. \ref{fig4}(a)-(c), we plot the filling factor, $\rm OS$, and $\rm DWO$ as a function of time. The parameters are chosen such that the lowest energy state in the unit-filled manifold of $\hat{H}_{\rm 0}$ is in the Haldane state $|\rm HI\rangle$. The frequency $\omega_d$ is chosen to be resonant with the transition between the vacuum state and $|\rm HI\rangle$, i.e. $\omega_d=\Delta E^L_{\rm HI}/L$. Note that this frequency is different than the one used in the previous section. With these conditions, we observe the hidden order in the transient dynamics emerging from the vacuum. This order eventually dies out at the steady state due to photon loss. We observe that the maximum value of the hidden order is increased with the amplitude of the drive. However the duration that the hidden order exists is reduced for a stronger drive. We found that the optimal value of the driving amplitude is around $1.25J-1.5J$, where the maximum $\rm OS$ is $\sim 0.05$ and the existence duration is $\sim 1/\Omega$.

In Fig. \ref{fig4}(d)-(f) we study the effect of photon loss. We find that the maximum $\rm OS$ is reduced when $\gamma$ is increased as expected. In Fig. \ref{fig4}(e)-(i), we study the effect of the $V$-term. We found that even when $V=0$, a transient SO order stills exists. This implies that the mechanism that generates SO is fundamentally different in the equilibrium case. However, when $V=0$ we also observe a large transient DWO whose magnitude is larger than SO. This implies that there is no hidden order during the evolution \cite{kennedy1992}. As $V$ is increased, this DWO is strongly suppressed while the SO remains appreciable leading to the transient hidden order.

We note that during the time evolution, the system is very far from the equilibrium $|\rm HI\rangle$. The $|\rm HI\rangle$ is an insulating state implying that the number of particles is conserved. In our situation, the number of particles is not conserved due the coherent drive and losses. Nevertheless, the dynamics of the driven-dissipative system as measured by the non-local hidden order significantly depends on the underlying equilibrium phase because the transient hidden order deceases when $\hat{H}^{\rm R}_{\rm 0}$ is far away from the $\rm HI$ phase. 


\section{Comparison with on-site Coherent Drive}
\label{sec:onsite}

Although the nearest-neighbor parametric driving discussed so far has been experimentally realized \cite{devoret2013}. It is not a common drive used in quantum optics. Previous literatures instead consider a more conventional one-photon drive \cite{carusotto2009, hartmann2010, carusotto2012, dimitris2012, jin2013, PhysRevA.87.053846,houck2016,PhysRevA.93.023821}
\begin{equation}
\hat{H}_{\rm local}=\Omega\sum_{i=1}^L(\hat{a}_ie^{i\omega_dt}+\hat{a}^\dagger_ie^{-i\omega_dt}).
\label{eq:os}
\end{equation}
In this section we will show that this drive has a different symmetry than the two-photon drive discussed in the previous section. To see this, let us consider the rotating frame defined by $\hat{R}$ as before. In this frame the drive becomes
\begin{equation}
\hat{H}_{\rm local}^{\text R}=-\omega_p \sum_{i}^{L}\hat{n}_i+\Omega\sum_{i=1}^L(\hat{a}_i+\hat{a}^\dagger_i).
\label{eq:os}
\end{equation}
Assuming a weak drive and mapping the system to a spin-chain system, the total Hamiltonian becomes
\begin{align}
\hat{H}^{S}_{\rm tot, loc}= &\sum_{i=1}^{L-1}\left(J\hat{S}^x_i\hat{S}^x_{i+1}+J\hat{S}^y_i\hat{S}^y_{i+1}+V \hat{S}^z_i\hat{S}^z_{i+1} \right) \nonumber \\
                           &+\sum_{i=1}^L\left(\frac{U}{2}(\hat{S}^z_i)^2 +\Omega\hat{S}^x_i\right).
\label{eq:hsos}
\end{align}
Again the term $\sum_i\hat{S}^z_i$ is dropped due to the unit-filling condition. We can see that the term $\Omega\hat{S}^x_i$ is not invariant under the transformation $\hat{S}^x_i\to-\hat{S}^x_i$. Hence it breaks the global $D_2$ symmetry. When applying the non-local unitary transformation $\hat{U}_{\rm KT}$, the terms becomes
\begin{equation}
\hat{U}_{\rm KT}\left(\sum_{i=1}^L\hat{S}^x_i\right)\hat{U}^{-1}_{\rm KT} = \sum_{i=1}^{L}\hat{S}_i^x\exp\left(i\pi\sum_{k=i+1}^L\hat{S}^x_k\right),
\end{equation}
which is highly non-local. Hence the FM phase in the transformed picture will be destroyed even for a weak drive. In the original picture, this means that $|\rm HI\rangle$ and its string order is destroyed in the presence of the on-site drive. This is confirmed by DMRG calculations, shown in Fig. \ref{fig3}. When performing the time evolution including dissipation using TEBD calculations, we also found that the SO remains zero throughout the time evolution.


\section{Conclusion}
\label{sec:conclusion}

We have shown that the dynamics of quantum many-body system driven by parametric process can exhibit hidden order which goes beyond local order parameters. The hidden order can arises in a transient case even when symmetry-breaking dissipation is included. We show that this drive respects the symmetry of the $\rm HI$ phase while the conventional on-site drive does not. Our work opens a new direction to explore the role of the non-local order and symmetry in non-equilibrium settings as well as its connection to the equilibrium SPT phases.

\textit{Acknowledgements.---} The authors acknowledge fruitful discussions with P. Nang Ma. The authors gratefully acknowledge financial support through the National Research Foundation and Ministry of Education Singapore (partly through the Tier 3 Grant ``Random  numbers  from  quantum  processes''). The research leading to these results has received funding from the European Research Council under the European Union’s Seventh Framework Programme (FP7/2007-2013) Grant Agreement No. 319286 Q-MAC, UK EPSRC funding EP/K038311/1, EPSRC under grant No. EP/P025110/1, EP/P009565/1, and EP/P01058X/1. Polisimulator project co-financed by Greece and the EU Regional Development Fund%



\begin{widetext}

\section*{Appendix}

\subsection{the hidden order}
\label{sec:the_hidden_string_order}
The hidden order was first introduced as a non-local order parameter to differentiate the Haldane spin-1 phase from other topologically-trivial spin phases \cite{nijs1989}. This later motivated the notion of SPT phases, which lie outside the conventional paradigm of Landau's symmetry breaking theory and so cannot be identified by a local order parameter \cite{cirac2008}. The hidden order was then generalized to the bosonic system by Ref.\cite{torre2006}. This bosonic hidden order was originally used to identify ground-state phases of the EBH model with unit filling $\bar{n}=1$ \cite{torre2006, berg2008, rossini2012}. It distinguishes the topological Haldane insulator (HI) phase ($\mathcal{O_{\rm S}}>0, \mathcal{O_{\rm DW}}=0$) from other topologically-trivial insulating phases which are the Mott phase ($\mathcal{O_{\rm S}}=\mathcal{O_{\rm DW}}=0$) and the density-wave phase ($\mathcal{O_{\rm DW}}>\mathcal{O_{\rm S}}>0$). The model also exhibits a superfluid phase when the interactions $U$ and $V$ are much smaller than the hopping strength $J$.

To visualize the structure of this hidden order, it is helpful to map the bosonic system to its equivalent spin-1 chain model with the total magnetization along the Z-axis fixed to zero \cite{torre2006}. This is done by truncating the bosonic Hilbert space up to n=0 photons per site. This is justified by the large on-site interaction $U\gg J$ required for the insulating phases. As a result, the bosonic Fock states  $\{|0\rangle_f, |1\rangle_f, |2\rangle_f\}$ can replaced by the spin-1 states $\{|+\rangle_s, |0\rangle_s, |-\rangle_s\}$ , i.e.  
\begin{equation}
|0\rangle_f \to |+\rangle_s, |1\rangle_f\to |0\rangle_s, \text{  and  } |2\rangle_f\to|-\rangle_s.
\end{equation}
In this picture, the Mott insulator, $|1111...\rangle_f$, and the density-wave, $|2020...\rangle_f$, become the ferromagnetic phase, $|0000...\rangle_s$, and the antiferromagnetic phase $|-+-+...\rangle_s$, respectively. The HI phase becomes the phase similar to the antiferromagnetic phase but with an arbitrary number of $|0\rangle_s$ between the states $|+\rangle_s$ and $|-\rangle_s$, e.g. $|+00-0+0000-...\rangle_s$ \cite{kennedy1992, Oshikawa1992, oshikawa2012} . Since the number of $|0\rangle_s$ between two spins is random, the two spins are uncorrelated, i.e. $\mathcal{O}_{\rm DW}=0$. However, the alternating long-range pattern between $|+\rangle_s$ and $|-\rangle_s$ is picked up by SO. 

To understand symmetry of the system, let's replace the bosonic operators with spin-one operators, i.e., $\hat{a}_i\to \hat{S}^+_i/\sqrt{2}$ and $\hat{n}_i\to \hat{\mathbbold{1}} -\hat{S}^z_j$. The extended Bose-Hubbard model with the unit-filling is then mapped to the effective spin Hamiltonian,
\begin{align}
\hat{H}^{S}_{0}= & J\sum_{i=1}^{L-1}\left(\hat{S}^x_i\hat{S}^x_{i+1}+\hat{S}^y_i\hat{S}^y_{i+1} \right) \nonumber \\
                           &+V \sum_{i=1}^{L-1}\hat{S}^z_i\hat{S}^z_{i+1}+\frac{U}{2}\sum_{i=1}^L(\hat{S}^z_i)^2.
\label{eq:hs}
\end{align}
The term $\sum_i\hat{S}^z_i$ is dropped, as it is zero for the unit-filled state. Similar to the EBH model, this model has gapped ground-state phases including all spin phases mentioned above \cite{kennedy1992, oshikawa2012}. It has the global $D_2=Z_2\times Z_2$ symmetry, i.e. $\pi-$rotation of all spins about $X$, $Y$, and $Z$ axes. The Haldane phase is a SPT phase protected by this symmetry, meaning that its edge states are robust against any perturbations that are smaller that the excitation gap and do not break the symmetry.


\subsection{Implementation of parametric pumping using circuit QED }
\label{app:circuitQED}

In this section, we propose an implementation of the Bose-Hubbard model driven by parametric pumping, using circuit-QED architecture. The cross-Kerr nonlinearity term $Vn_in_j$ has already been discussed in the literature \cite{Jin,Holland} and can be integrated to our circuit. So we will not reproduce it here. Note that this term has also been implemented experimentally for a dimer \cite{vterm}. Nevertheless, circuit designs discussed in \cite{Jin,Holland,vterm} do lead to extra terms in the Hamiltonian that needed further investigation.

Our circuit diagram is shown in Fig.\ref{fig:circuit-QED}. The flux variable is defined as $\phi_i=-\int V_i dt$, where $V_i$ is a voltage at the corresponding position. As will be shown below, this quantity can be quantised to the form $\phi_i=\alpha (a_i+a^\dagger_i)$, where $a_i, a_i^\dagger$ are bosonic operators of an `artificial'' photon at site $i$ and $\alpha$ is some constant that depends on the circuit's elements. We first describe the rules of Josephson junctions which introduce various kinds of nonlinearities to the system and then explicitly show how to quantise the circuit.

The first Josephson junction $E_{J,U}$ (labelled in orange in Fig.\ref{fig:circuit-QED}) corresponds to a $\chi^{(3)}$ nonlinear material, which gives rise to the on-site Kerr nonlinearity $\frac{U}{2}n_i(n_i-1)$. The junction is biased by the magnetic flux $\Phi_g=\pi\phi_0$, where $\phi_0=\hbar/2e$, and shunted by a small inductor $L'$ to produce a repulsive interaction $U>0$. The second Josephson junction $E_{J,\Omega_p}$ (labelled in yellow in Fig.\ref{fig:circuit-QED}) corresponds to a $\chi^{(2)}$ nonlinear material, which responsible for a parametric-down-conversion (PDC) process. The PDC process converts a pumped photon with frequency $2\omega_p$ into a pair of photons with frequency $\omega_p$. Here the pumped photons come from an oscillating flux bias $\Phi_b (t) = \pi \phi_0 /2 + \phi_b(t)$. 

As discussed in \cite{Bardyn}, this PDC process leads to both nearest-neighbour parametric pumping of the form $(a^\dagger_ia^\dagger_{i+1}+h.c.)$ and on-site parametric pumping of the form $(a^{\dagger 2}_i+a_i^2)$. The latter can be eliminated by introducing an extra on-site PDC process (labelled in a dotted box in Fig.\ref{fig:circuit-QED}). This extra component is driven by a coherent voltage source $\psi$, whose phase differs from that of $\phi_b(t)$ by $\pi$.

We now show how to quantise the the circuit by following the standard procedure \cite{Devoret_lecture}. We first write down the circuit's Lagrangian as $\mathcal{L}=\sum_i(\mathcal{L}^{\text{on-site}}_i+\mathcal{L}^{\text{hopping}}_i+\mathcal{L}_i^{pump}+\mathcal{L}^{\text{onsite-PDC}}_i)$ where
    \begin{align}
    \mathcal{L}^{\text{on-site}}_i  = &\frac{1}{2}C_J\dot{\phi}_i^2-\frac{1}{2L'}\phi_i^2+E_{J,U}\cos\left(\frac{\phi_i+\pi\phi_0}{\phi_0}\right), \label{eq:circuit_kerr}  \\ 
    \mathcal{L}^{\text{hopping}}_i  = & \frac{1}{2}C(\dot{\phi}_i-\dot{\phi}_{i+1})^2-\frac{1}{2L}(\phi_i-\phi_{i+1})^2, \\ 
    \mathcal{L}^{\text{pump}}_i  = & E_{J,\Omega_p}\cos\left(\frac{\phi_i-\phi_{i+1}+ \pi\phi_0/2+\phi_b(t)}{\phi_0}\right), \label{eq:circuit_drive} \\ 
    \mathcal{L}^{\text{onsite-PDC}} = & \frac{1}{2}C(\dot{\phi}_i-\dot{\psi})^2 +E_{J,\Omega_p}\cos\left(\frac{\phi_i-\psi+\pi\phi_0/2}{\phi_0}\right) \label{eq:pdc}
    \end{align}
Assuming $C/(C_J+3C)\ll 1$, the Hamiltonian can then be obtained using the Legendre transformation \cite{Nunnenkamp}. A conjugate momentum of $\phi_i$ is defined as $q_i=\sqrt{3C+C_J}\partial \mathcal{L}/\partial \dot{\phi_i}$. Both $\phi_i$ and $q_i$ are then quantised by defining ladder operators $a_i$, $a^\dagger_i$ according to $\phi_i  = (\tilde{L}/4\tilde{C})^{1/4}(a_i+a^\dagger_i)$ and $q_i = i(\tilde{C}/4\tilde{L})^{1/4}(-a_i+a^\dagger_i)$, where $\tilde{C}= C_J+3C$ and $\tilde{L}=\left[1/L' +3/L - E_{J,U}/\phi_0^2\right]^{-1}$ are effective capacitance and effective inductance, respectively. It follows that $[a_i,a^\dagger_j]=\delta_{ij}$. In addition, after the Legendre transformation, the quadratic terms in $\mathcal{L}$ are transformed into $\sum_i\omega_ca^\dagger_ia_i$, where $\omega_c=1/\sqrt{\tilde{L}\tilde{C}}$ is a frequency of the artificial photon. We can see that by adding a small shunting inductor $L'$, $\omega_c$ is guaranteed to be real.

\begin{figure}
    \includegraphics[width=8.5cm,height=4.5cm]{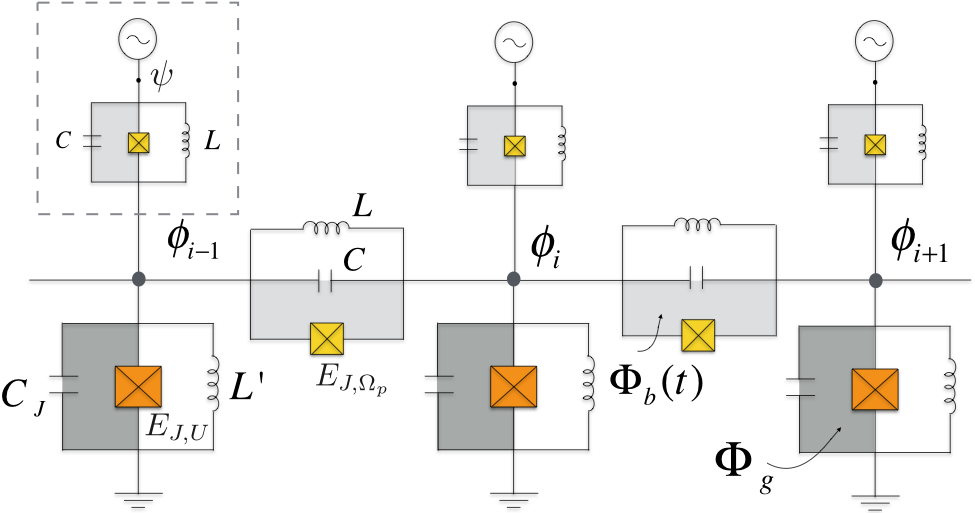}
    \caption{Proposed circuit diagram that implements the Bose-Hubbard model and parametric driving.}
    \label{fig:circuit-QED}
\end{figure}

To see the emergence of the on-site Kerr nonlinearity $U$ and PDC, we first notice that the Legendre transformation only introduces a minus sign to the `potential'' terms, including all the cosine terms in $\mathcal{L}$. Hence, the onsite-Kerr nonlinearity can be seen by expanding the cosine function in Eq.(\ref{eq:circuit_kerr}), taking into account the normal ordering as \cite{Leib}
    \begin{align}
    E_{J,U} &\cos  \left( \lambda(a_i+a^\dagger_i)\right) \nonumber = E_{J,U}e^{-\lambda^2/2}\left(1-\lambda^2a_i^\dagger a_i + \frac{\lambda^4}{4}a_i^\dagger a_i^\dagger a_i a_i+...\right).
    \label{eq:cosine_expansion}
    \end{align}
where $\lambda =(2E_{\tilde{C}}/E_{\tilde{L}})^{1/4}$, with $E_{\tilde{C}}=e^2/2\tilde{C}$ and $E_{\tilde{L}}=\phi_0^2/\tilde{L}$. For a large $E_{\tilde{L}}/E_{\tilde{C}}$, we can neglect the terms that are higher than the forth order \cite{Koch}.

The parametric pumping term comes directly from the sine expansion in Eq.(\ref{eq:circuit_drive}). For illustrative purpose, we neglect the normal-ordering and consider the sine expansion up to the third order as

    \begin{align}
    E_{J,\Omega_p} & \sin \left( \frac{\phi_i-\phi_{i+1}+\phi_b}{\phi_0}\right) \nonumber \approx 
\frac{E_{J,\Omega_p}}{\phi_0}\left( \phi_i-\phi_{i+1}\right)-\frac{E_{J,\Omega_p}}{3!\phi_0^3}\left( \phi_i-\phi_{i+1}+\phi_b\right) ^3,
    \end{align}
where we neglect the term $E_{JJ}\phi_b/\phi_0$, since it does not act on the system. The linear term can  also be eliminated by applying a current bias $I$ at both ends of the array. After rotating wave approximation, the only third-order terms that survive are of the forms ($b^\dagger a_ia_{i+1}+h.c.$) and $(b^\dagger a_i^2+b^\dagger a_{i+1}^2+h.c.)$, where $b^\dagger$ is a creator of the field $\phi_b$. The latter is cancelled by the onsite PDC process in Eq.\ref{eq:pdc}.
    
Finally by explicitly writing down the time dependence of $b$ and $b^\dagger$ and replacing them with $c-$numbers, the Hamiltonian can be cast into the form
    \begin{align}
    \mathcal{H}= \sum_i(\omega+\delta\omega)a^\dagger_ia_i+\frac{U}{2}a^\dagger_ia^\dagger_ia_ia_i-J(a^\dagger_ia_{i+1}+h.c.) +\Omega_p(e^{i2\omega_pt}a_ia_{i+1}+h.c.),
    \end{align}
where $\omega = 1/\sqrt{\tilde{L}\tilde{C}}$, $\delta \omega = E_J\lambda^2(1-e^{-\lambda^2/2})$, $U/2 = E_J\lambda^4e^{-\lambda^2/2}/4$ and $J = \omega /2(\tilde{L}/L-C/\tilde{C})$. The parametric pumping coefficient $\Omega_p$ is directly proportional to $E_{JJ}$. However, its explicit form depends on the relation between $\phi_b$ and $(b+b^\dagger)$ and hence depends on how $\phi_b$ is generated.

Our circuit allows the Hamiltonian parameters to be tuned independently: $\mu$ can be tuned directly by changing $\omega_p$, $U$ comes from the first Josephson junction $E_{J,U}$, $J$ comes from the coupling $LC$ oscillator while $\Omega_p$ independently comes from the second Josephson junction $E_{J,\Omega_p}$. As an example, $U/J\sim 10$ can be realistically obtained by using  $\tilde{L}/L\sim 5\times 10^{-3}$, $\lambda \sim 0.4$ and $E_J/E_C\sim 10^5$\cite{Devoret,You}. For this setting, we would have a negligible frequency correction $\delta \omega/\omega\sim 0.02$. Noted that this value of $\lambda$ also ensures that it is a good approximation to expand the cosine term in Eq.(\ref{eq:cosine_expansion}) up to the fourth order, as for the case of a transmon qubit \cite{Koch}.

\end{widetext}
 
\bibliography{ref,ref1}

\end{document}